\documentclass{article}%
\usepackage{amsfonts}
\usepackage{amsmath}%
\usepackage{amssymb}%
\usepackage{graphicx}
\providecommand{\U}[1]{\protect\rule{.1in}{.1in}}
\newtheorem{theorem}{Theorem}[section]

\newtheorem{corollary}[theorem]{Corollary}

\newtheorem{remark}[theorem]{Remark}

\newenvironment{MSC 2020}[1][MSC 2020]{\noindent \textbf{#1} }{}
\newenvironment{keywords}[1][Keywords]{\noindent\textbf{#1} }{}

\begin{document}
\date{}
\title{Characterization of  generalized quasi-Einstein manifolds and modified gravity }
\author{Uday Chand DE$^{1} $ and Hülya BAĞDATLI YILMAZ $^{2}$\\ \\
	$ ^{1} $ Calcutta University, 	Department of Pure Mathematics,\\ West Bengal/India\\ 
	\\$ ^{2} $Marmara University, Faculty of Sciences,\\ Department of Mathematics, \\ Istanbul/Turkey \\  \\
	 E-mails: uc\_de@yahoo.com$^{1} $; hbagdatli@marmara.edu.tr$^{2} $\\
 }
\maketitle

\begin{abstract}
In this work, a detailed examination of a specific case of a generalized quasi-Einstein manifold $ (GQE)_{n} $ is provided. We begin by exploring generalized quasi-Einstein spacetimes under certain conditions, with a primary focus on cases that admit a parallel time-like vector field. Among the findings, it is demonstrated that such spacetimes can be categorized as generalized Robertson-Walker, Robertson-Walker, and quasi-constant curvature spacetimes. Additionally, the physical implications of these results are discussed. Furthermore, $ (GQE)_{4} $ spacetimes that admit $\mathcal{F}(\mathfrak{R})$-gravity and feature a parallel unit time-like vector field are investigated. In this context, a flat Friedmann-Robertson-Walker (FRW) model is established to extract exact cosmological solutions for the Hubble parameter and the scale factor. Various energy conditions are analyzed based on geometric constraints arising from $\mathcal{F}(\mathfrak{R})$-gravity. Finally, a non-trivial 4-dimensional GQE spacetime is constructed and its exact, closed-form potential function is obtained to verify the analytical feasibility of our model.
\end{abstract}

\begin{MSC 2020} 53B30; 53C25; 53C50; 53Z05
\end{MSC 2020}


\section{Introduction}
Complete Riemannian manifolds $(\mathcal{M}^{n}, \mathfrak{g})$, where $n \geq 3$, are classified as generalized quasi-Einstein manifolds if they include smooth functions $\mu$, $\lambda$, and $\mathfrak{f}$ that are defined on $\mathcal{M}$ and satisfy the following equation:

\begin{equation}
Ric + \nabla^{2}\mathfrak{f} - \mu \, d\mathfrak{f} \otimes d\mathfrak{f} - \lambda \mathfrak{g} = 0. \label{eq1}
\end{equation}
In this equation, $Ric $ denotes the Ricci tensor, while  $\nabla^{2}\mathfrak{f} $ represents the Hessian of the smooth function  $\mathfrak{f} $. In \cite{bibC2012}, it is established that such a manifold, which is defined by a harmonic Weyl tensor and a radial Weyl curvature that vanishes, can be locally represented as a special warped product.

In the literature, there are two different notions of generalized quasi-Einstein manifolds defined as follows:

According to Chaki \cite{bibC2001}, a non-flat Riemannian manifold $(\mathcal{M}^n, g)$ $(n \ge 3)$ is called a generalized quasi-Einstein manifold if its Ricci tensor $Ric$ of type $(0,2)$ is non-zero and satisfies the condition
\begin{equation*}
	Ric(X,Y)=a g(X,Y)+b A(X)A(Y)+c\ \left[A(X)B(Y)+  A(Y)B(X)\right] ,
\end{equation*}
where $a, b, c$ are non-zero scalar functions, and $A, B$ are 1-forms associated with mutually orthogonal unit vector fields. In \cite{bibGD2016}, utilizing this specific Chaki framework, the geometric and physical properties of generalized quasi-Einstein spacetimes were investigated under certain conditions.

On the other hand, De and Ghosh \cite{bibDG2004} introduced an alternative geometric structure under the same name. They defined a generalized quasi-Einstein manifold as a non-flat Riemannian manifold whose Ricci tensor satisfies
\begin{equation*}
	Ric(X,Y)= ag(X,Y)+b A(X)A(Y)+cB(X) B(Y),
\end{equation*}
where $a, b, c$ are non-zero scalar functions, and $A, B$ are 1-forms corresponding to mutually orthogonal unit vector fields.

Throughout this paper, we focus on a different generalized framework that extends these classical definitions, as expressed in (\ref{eq1}).

Several interesting geometric structures can be seen as natural examples of quasi-Einstein manifolds. These include Einstein manifolds, in which both $ \mathfrak{f} $ and $ \lambda $ are constants, and gradient Ricci solitons that exhibit a constant $ \lambda $ with $ \mu = 0 $. Additionally, gradient Ricci almost solitons are considered in cases where $ \mu = 0$ \cite{bibPRS10}. Quasi-Einstein manifolds arise when both $ \mu $ and $ \lambda $ are constants \cite{bibCSW2011, bibCMMR13}. 

The $m$-quasi Einstein metric is introduced by Case et al. \cite{bibCSW2011}. In this case, $ \mu=\frac{1}{m} $ for a positive integer $ m , (0<m\leqslant+\infty )$ and $ \lambda\in \mathbb{R} $. The authors in \cite{bibBR2014} and \cite{bibD2015} delved into a specific case of this result. When $ m = 2 $, the $m$-quasi Einstein metric is recognized as a vacuum near-horizon geometry \cite{bibG25}, and (\ref{eq1}) reduces to the vacuum near-horizon geometry equation of spacetime with $ \lambda $ playing the role of cosmological constant. 

In this work, we aim to explore a particular case of (\ref{eq1}). Specifically, we will use the following definition:

A complete Riemannian manifold $(\mathcal{M}^{n}, \mathfrak{g})$, defined for dimensions $n \geq 3$, is classified as a generalized quasi-Einstein manifold, denoted as $(GQE)_{n}$, if there are non-zero constants $\mu$ and $\lambda$ present, along with a smooth function $\mathfrak{f}$ defined on $\mathcal{M}$ that fulfills 
\begin{equation*}
	Ric + \nabla^{2}\mathfrak{f} - \mu \, d\mathfrak{f} \otimes d\mathfrak{f} - \lambda \mathfrak{g} = 0.
\end{equation*}

In the local coordinate system, it becomes
\begin{equation}
\mathfrak{R}_{ij} + \mathfrak{f}_{ij} - \mu \,  \mathfrak{f}_{i}  \mathfrak{f}_{j} - \lambda \mathfrak{g}_{ij} = 0,\label{eq2}
\end{equation}
where $  \mathfrak{f}_{i}=\nabla_{i}\mathfrak{f} $, $  \mathfrak{f}_{ij}=\nabla_{j}\nabla_{i}\mathfrak{f} $, and $ \mathfrak{R}_{ij} $ denotes the components of the Ricci tensor $ Ric $.

A Lorentzian manifold of dimension $n$ is a special class of semi-Riemannian manifold, which holds significant relevance in the domains of cosmology and relativity due to its capacity to classify vectors based on their causal characteristics. A Lorentzian manifold that possesses a globally time-like vector field is categorized as a spacetime.

In the study of spacetime, a perfect fluid (PF) spacetime is characterized by the condition that the Ricci tensor is non-zero and adheres to the following equation:
\begin{equation}
\mathfrak{R}_{ij} = \eta \mathfrak{g}_{ij} + \tau \mathfrak{u}_{i} \mathfrak{u}_{j}, \label{eq3}
\end{equation}
where $ \eta $ and $ \tau $ are scalars. In this equation, $ \mathfrak{u} $ represents a unit time-like vector, meaning that $ \mathfrak{u}_{i} \mathfrak{u}^{i} = -1 $ \cite{bib17, 36}.

Recently, researchers have explored geometric flows in the quest to identify a cosmological model that behaves like a PF spacetime. Blaga \cite{bibB2020} examined regular and Einstein solitons within a PF spacetime framework. Meanwhile, Venkatesha and Kumara \cite{bibVK19} investigated regular solutions for perfect fluid that follows a torse forming vector field.

A generalized Robertson-Walker (GRW) spacetime is defined as an $ n $- dimensional Lorentzian manifold, characterized by the following metric, which serves as a fundamental framework for understanding the geometric structure of space-time:

\begin{equation*}
	d\mathfrak{s}^2 = -d\mathfrak{t}^2 +\mathfrak{a}^2(\mathfrak{t}) \tilde{\mathfrak{g}}_{ij}(\vec{\mathfrak{x}}) d\mathfrak{x}^i d\mathfrak{x}^j,
\end{equation*}
here $ \tilde{\mathfrak{g}}_{ij}(\vec{\mathfrak{x}}) $ represents the metric tensor of a Riemannian submanifold and $\mathfrak{a} > 0$ is a scale function or smooth warping. This structure provides a broader framework than the Robertson-Walker (RW) spacetime, making it essential for the study of large-scale cosmology. Both GRW and RW spacetimes play crucial roles in cosmology by effectively illustrating the temporal evolution of a homogeneous and isotropic universe \cite{bibA1995, bib17, bib031, bib2017, 16, bib98}.

The theory of general relativity provides a comprehensive framework for understanding the distribution of matter within spacetime through the energy-momentum tensor (EMT), denoted as \(\mathfrak{T}_{ij}\). The geometric structure of spacetime is profoundly influenced by the Ricci tensor, which is interconnected with the EMT via Einstein's field equations (EFEs).

 The characteristics of PF are defined by its rest frame mass density and isotropic pressure. Consequently, the EMT $\mathfrak{T}_{ij}$ in a PF spacetime can be formulated as follows:
\begin{equation}
\mathfrak{T}_{ij} =\mathfrak{p} \mathfrak{g}_{ij} + (\mathfrak{p} + \sigma)  \mathfrak{u}_{i}  \mathfrak{u}_{j}. \label{eq4}
\end{equation}
In the foregoing equation, let $\mathfrak{p} $ denote the isotropic pressure, while $\sigma $ signifies the energy density. A perfect fluid is classified as isentropic if it depends on a barotropic equation of state (EoS) $\mathfrak{p} = \mathfrak{p}(\sigma)$. Several specific cases can further delineate the nature of this spacetime:  $ \sigma = 3\mathfrak{p} $ corresponds to the radiation era;  $\mathfrak{p} = \sigma $ characterizes a stiff matter fluid; $\mathfrak{p} + \sigma = 0 $ is associated with the dark energy era \cite{bibDS1999, bib25}.

The EFEs, excluding the cosmological constant, are given by:
\begin{equation}
\mathfrak{R}_{ij} - \frac{\mathfrak{R}}{2} \mathfrak{g}_{ij} = \kappa \mathfrak{T}_{ij}, \label{eq5}
\end{equation}
in which \(\kappa\) indicates the gravitational constant and $\mathfrak{R}$ represents the scalar curvature, $\mathfrak{R}=\mathfrak{R}_{ij}\mathfrak{g}^{ij}$. In the field of cosmology, the scalar curvature plays a pivotal role in elucidating the structure of the universe. Under conditions of vacuum, the EFEs can be simplified to $\mathfrak{R}_{ij} = 0$. This represents a non-linear differential equation, which leads to the conclusion that the scalar curvature $ \mathfrak{R}  $ is also zero.

Utilizing (\ref{eq3}) and (\ref{eq4}) in (\ref{eq5}), we deduce that
$$\eta= \frac{\kappa(\sigma-\mathfrak{p})}{n-2} \quad \text{and} \quad \tau=\kappa(\sigma+\mathfrak{p}),$$
Here, we have used the relation $\mathfrak{R}=n\eta-\tau$ , which is obtained by contracting (\ref{eq3}) by $ \mathfrak{g}^{ij}$.

The general relativity theory is fundamentally based on a metric framework. In contrast, $\mathcal{F}(\mathfrak{R})$-gravity theories allow for the characterization of spacetime through both metric and affine structures. This distinction gives rise to two primary approaches within $\mathcal{F}(\mathfrak{R})$-gravity theories: the Palatini formalism, which focuses on the affine structure, and the metric formalism. Moreover, $\mathcal{F}(\mathfrak{R})$-gravity theories can be examined through scalar-tensor gravity theories, such as the Brans–Dicke theory, which emphasizes the role of scalar fields in gravitational interactions \cite{bibBD1961}. From a cosmological perspective, the necessity of these theoretical frameworks becomes evident when addressing the universe's accelerated expansion and the enigmatic nature of dark energy, which remain among the most challenging problems in modern cosmology. To address these issues, $\mathcal{F}(\mathfrak{R})$-gravity provides a robust alternative to the standard cosmological model by modifying the geometric part of the Einstein-Hilbert action. By utilizing these modified frameworks along with perfect fluid models, one can gain deeper insights into the behavior of cosmic matter under extreme gravitational fields. Under this scope, the geometric properties established by the Weyl tensor and its divergence play a pivotal role in characterizing the underlying curvature profile.

Accordingly, the (0,4) Weyl tensor $\mathcal{C}_{hijk}$ on an  $n$-dimensional semi-Riemannian manifold is defined as follows: 
\begin{equation}
	\mathcal{C}_{hijk}=\mathfrak{R}_{hijk}-\frac{1}{n-2}\left\lbrace \mathfrak{R}_{hk}\mathfrak{g}_{ij}-\mathfrak{R}_{hj}\mathfrak{g}_{ik}+\mathfrak{R}_{ij}\mathfrak{g}_{hk}-\mathfrak{R}_{ik}\mathfrak{g}_{hj}\right\rbrace \label{eq6}
\end{equation}
\begin{equation*}
	+\frac{\mathfrak{R}}{(n-1)(n-2)}\left\lbrace \mathfrak{g}_{hk}\mathfrak{g}_{ij}-\mathfrak{g}_{hj}\mathfrak{g}_{ik}\right\rbrace, 
\end{equation*}
where $\mathfrak{R}_{jkl}^{i}$ denotes the curvature tensor of type  $(1,3)$ and $\mathfrak{R}_{hjkl}=\mathfrak{g}_{ih}\mathfrak{R}_{jkl}^{i}. $

The divergence of the Weyl tensor is subsequently expressed as \cite{bibP2001}:
\begin{equation}
 \nabla_{h}\mathcal{C}^{h}_{ijk} = \frac{n-3}{n-2}\left[ \nabla_{j}\mathfrak{R}_{ik}-\nabla_{i}\mathfrak{R}_{jk}-\frac{1}{2(n-1)}\left( \mathfrak{g}_{ik}\nabla_{j}\mathfrak{R}-\mathfrak{g}_{jk}\nabla_{i}\mathfrak{R}\right) \right]. \label{eq06}
\end{equation}

In this framework, the foundation of our study lies in modified $\mathcal{F}(\mathfrak{R})$-gravity theories, which integrate higher-order curvature terms into the classical Einstein-Hilbert Lagrangian density \cite{bibS1980}. Originally designed to address quantum effects that may arise in the early universe, these frameworks incorporating geometric correction terms are commonly referred to as nonlinear, variational, or higher-order gravitational theories \cite{bibCRM1998, bibCRM98, bibS1998}.

 Crucially, rather than limiting the analysis to purely algebraic characterizations of $(GQE)_4$ spacetimes, this work bridges the gap between abstract tensor constraints and dynamic spacetime evolutions by mapping these geometric fields onto a realistic cosmological setup. By implementing a flat Friedmann-Robertson-Walker (FRW) metric directly onto the modified field equations of $\mathcal{F}(\mathfrak{R})$-gravity, we can extract explicit, time-dependent exact solutions for the Hubble parameter and the scale factor. Consequently, analyzing the gravitational fields of generalized quasi-Einstein spacetimes within these rigorous geometric bounds allows for a deeper exploration of cosmic matter distributions. In this context, evaluating the classical energy conditions, namely, the weak, dominant, and strong energy conditions, serves as a crucial physical filter to test the viability of the derived cosmological models, ensuring they delineate realistic matter configurations rather than unphysical states.

The subsequent sections of this work are organized as follows: Section 2 commences with an examination of the geometric characteristics of $(GQE)_{n}$ spacetimes under certain well-defined conditions, shifting later to the analysis of those admitting a parallel unit time-like vector field. Section 3 investigates $(GQE)_{4}$ spacetimes endowed with a parallel unit time-like vector field within the framework of $\mathcal{F}(\mathfrak{R})$-gravity. In Section 4, we establish a flat Friedmann-Robertson-Walker (FRW) cosmological model and perform a detailed Hubble parameter analysis to extract exact time-dependent power-law solutions. Section 5 presents a detailed analysis of various energy conditions based on the geometric constraints investigated throughout this study. Finally, in Section 6, we demonstrate the explicit geometric consistency and analytical feasibility of our theoretical results by constructing a non-trivial 4-dimensional GQE spacetime example with strict non-zero constants $\lambda$ and $\mu$, and obtaining its complete closed-form potential function.

\section{Generalized quasi-Einstein spacetimes  } 
Let's now deal with $(GQE)_{n}$ spacetimes.

Covariant differentiation of (\ref{eq2}) yields
	\begin{equation}
		\nabla_{l}\mathfrak{R}_{ij} +\nabla_{l}\mathfrak{f}_{ij}-\mu\left[ (\nabla_{l}\mathfrak{f}_{i})\mathfrak{f}_{j}+ \mathfrak{f}_{i}(\nabla_{l}\mathfrak{f}_{j})\right]=0.  \label{eq7}  
	\end{equation}
(\ref{eq7}) can be rearranged in the following form:

	\begin{equation}
	\nabla_{l}\mathfrak{f}_{ij}=-\nabla_{l}\mathfrak{R}_{ij} +\mu\left[ (\nabla_{l}\mathfrak{f}_{i})\mathfrak{f}_{j}+ \mathfrak{f}_{i}(\nabla_{l}\mathfrak{f}_{j})\right].  \label{eq8}  
	\end{equation}
By changing the indices $ j $ and $ l $ in (\ref{eq8}) and then subtracting the resulting equation from (\ref{eq8}), one can obtain

	\begin{equation}
\nabla_{l}\mathfrak{f}_{ij}-\nabla_{j}\mathfrak{f}_{il}=\nabla_{j}\mathfrak{R}_{il}-\nabla_{l}\mathfrak{R}_{ij} +\mu\left[ (\nabla_{l}\mathfrak{f}_{i})\mathfrak{f}_{j}- (\nabla_{j}\mathfrak{f}_{i})\mathfrak{f}_{l}\right]. \label{eq9}
	\end{equation}
Considering Ricci identity, (\ref{eq9}) becomes
		\begin{equation}
	\mathfrak{f}_{h}\mathfrak{R}_{ijl}^{h}=\nabla_{j}\mathfrak{R}_{il}-\nabla_{l}\mathfrak{R}_{ij} +\mu\left[ (\nabla_{l}\mathfrak{f}_{i})\mathfrak{f}_{j}- (\nabla_{j}\mathfrak{f}_{i})\mathfrak{f}_{l}\right]. \label{eq10}
	\end{equation} 
Suppose that this spacetime is a PF spacetime. Then, from (\ref{eq3}), we can reach 
	\begin{equation}
	\nabla_{l}\mathfrak{R}_{ij} = \eta_{l} \mathfrak{g}_{ij} + \tau_{l} \mathfrak{u}_{i} \mathfrak{u}_{j} +\tau \left[(\nabla_{l}\mathfrak{u}_{i})\mathfrak{u}_{j}+\mathfrak{u}_{i}(\nabla_{l}\mathfrak{u}_{j})\right] , \label{eq11}
	\end{equation}
where $\eta_{l}=\nabla_{l}\eta $ and $\tau_{l}=\nabla_{l}\tau $.

After changing the indices $ j $ and $ l $ in (\ref{eq11}) and  substituting the resulting equation and (\ref{eq11}) into (\ref{eq10}), we have
	\begin{equation*}
	\mathfrak{f}_{h}\mathfrak{R}_{ijl}^{h}=\eta_{j} \mathfrak{g}_{il}-\eta_{l} \mathfrak{g}_{ij}+\left[ \tau_{j} \mathfrak{u}_{l}-\tau_{l} \mathfrak{u}_{j}\right] \mathfrak{u}_{i}  
\end{equation*}
\begin{equation}
		+\tau\left[  (\nabla_{j}\mathfrak{u}_{i})\mathfrak{u}_{l} -(\nabla_{l}\mathfrak{u}_{i})\mathfrak{u}_{j}+\mathfrak{u}_{i}\left( (\nabla_{j}\mathfrak{u}_{l})-(\nabla_{l}\mathfrak{u}_{j})\right) \right]  \label{eq12}
\end{equation}
\begin{equation*}
	+\mu \left[ (\nabla_{l}\mathfrak{f}_{i})\mathfrak{f}_{j}- (\nabla_{j}\mathfrak{f}_{i})\mathfrak{f}_{l}\right].
\end{equation*}
From (\ref{eq2}), it follows that
\begin{equation}
	\mathfrak{f}_{ij}=-\mathfrak{R}_{ij} +  \mu \,  \mathfrak{f}_{i}  \mathfrak{f}_{j} + \lambda \mathfrak{g}_{ij}.\label{eq13}
\end{equation}
Multiplying (\ref{eq13}) by $\mathfrak{f}_{l}  $ gives
\begin{equation}
	\mathfrak{f}_{ij}\mathfrak{f}_{l}=-\mathfrak{R}_{ij}\mathfrak{f}_{l} +  \mu \,  \mathfrak{f}_{i}  \mathfrak{f}_{j}\mathfrak{f}_{l} + \lambda \mathfrak{g}_{ij}\mathfrak{f}_{l}.\label{eq14}
\end{equation}
By changing the indices $ j $ and $ l $ in (\ref{eq14}) and then subtracting the resulting equation from (\ref{eq14}),  we reveal that
\begin{equation}
	\mathfrak{f}_{ij}\mathfrak{f}_{l}-\mathfrak{f}_{il}\mathfrak{f}_{j}=\mathfrak{R}_{il}\mathfrak{f}_{j}-\mathfrak{R}_{ij}\mathfrak{f}_{l} + \lambda \left[ \mathfrak{g}_{ij}\mathfrak{f}_{l}-\mathfrak{g}_{il}\mathfrak{f}_{j}\right]. \label{eq15}
\end{equation}
	Multiplying (\ref{eq15}) by $\mathfrak{g}^{ij} $ and $\mathfrak{u}^{l} $, respectively, we can get
\begin{equation}
(\mathfrak{f}_{ij}\mathfrak{f}_{l}-\mathfrak{f}_{il}\mathfrak{f}_{j})\mathfrak{g}^{ij}\mathfrak{u}^{l}=\mathfrak{R}_{il}\mathfrak{f}^{i}\mathfrak{u}^{l} +  \left[\lambda(n-1)-\mathfrak{R}\right] \mathfrak{f}_{l}\mathfrak{u}^{l}.\label{eq16}
\end{equation}
 
On the other hand, multiplying (\ref{eq12}) by $\mathfrak{g}^{ij} $ and $\mathfrak{u}^{l} $, respectively, it can be found that
	\begin{equation}
\mathfrak{f}_{h}\mathfrak{R}_{l}^{h}\mathfrak{u}^{l}=(1-n)\eta_{l}\mathfrak{u}^{l}-\tau(\nabla_{j}\mathfrak{u}^{j})+\mu\left[\mathfrak{f}_{jl}\mathfrak{f}_{k}-\mathfrak{f}_{jk}\mathfrak{f}_{l}\right] \mathfrak{g}^{jk}\mathfrak{u}^{l}  .\label{eq17}
	\end{equation}
	Putting (\ref{eq16}) in (\ref{eq17}), we get
\begin{equation}
\mathfrak{f}_{h}\mathfrak{R}_{l}^{h}\mathfrak{u}^{l}=(1-n)\eta_{l}\mathfrak{u}^{l}-\tau(\nabla_{j}\mathfrak{u}^{j})+\mu\left[  \mathfrak{R}_{jl}\mathfrak{f}^{j}\mathfrak{u}^{l}+ \left( \lambda (n-1) -\mathfrak{R}\right) \mathfrak{f}_{l}\mathfrak{u}^{l} \right] .\label{eq18}
\end{equation}

Moreover,  multiplying (\ref{eq2}) by $ \mathfrak{f}^{j} $ and $ \mathfrak{u}^{l} $ gives
	\begin{equation}
	\mathcal{R}_{jl}\mathfrak{f}^{j}\mathfrak{u}^{l}=(\eta-\tau)\mathfrak{f}_{l}\mathfrak{u}_{l}. \label{eq19}
	\end{equation}

Let's assume that the scalar function $ \eta $ and the smooth function $\mathfrak{f}  $ satisfy $ \eta_{l}\mathfrak{u}^{l}=0  $ and $ \mathfrak{f}_{l}\mathfrak{u}^{l}=0  $. Then, considering (\ref{eq18}) and (\ref{eq19}), we achieve
	\begin{equation}
\tau(	\nabla_{j}\mathfrak{u}^{j})=0.\label{eq20}
	\end{equation}
(\ref{eq20}) means that $\tau=0  $ or $ \nabla_{j}\mathfrak{u}^{j}=0 $. Thus, if $ \nabla_{j}\mathfrak{u}^{j}=0 $, then the velocity vector is divergence-free, implying that the expansion scalar of the fluid vanishes \cite{bib25}. In the second case, if $\tau=0$, then (\ref{eq3}) implies that the spacetime is Einstein, and consequently, by (\ref{eq06}),  $ div \:\mathcal{C}=0$. Moreover, $\mathfrak{p}+\sigma=0 $, indicating that the spacetime corresponds to the dark energy era. In \cite{19}, it was demonstrated that a GRW spacetime is a PF spacetime if and only if $ div \:\mathcal{C}=0$.

	Therefore we can state the following result:
	\begin{theorem}\label{1}
	Let a  $ (GQE)_{n} $ spacetime be a PF spacetime in which the scalar function $ \eta $ and the smooth function $\mathfrak{f}  $ satisfy $ \eta_{l}\mathfrak{u}^{l}=0  $ and $ \mathfrak{f}_{l}\mathfrak{u}^{l}=0  $. Then, the spacetime either indicates the dark energy era and becomes a GRW spacetime, or the expansion scalar of the fluid may vanish.
	\end{theorem}

Let's now suppose that $ \varphi^{i} $ satisfies the condition
\begin{equation}
	\theta=\mathfrak{f}_{i}\varphi^{i}.\label{eq21}
\end{equation}
Taking $ \varphi $ as a parallel vector field, that is,
\begin{equation}
\nabla_{j}\varphi_{i}=0,\label{eq22}
\end{equation}
then it follows from (\ref{eq21}) and (\ref{eq22}) that
\begin{equation}
	\theta_{j}=\mathfrak{f}_{ij}\varphi^{i}.\label{eq23}
\end{equation}
Applying the covariant derivative on (\ref{eq23}) gives
\begin{equation}
	\nabla_{k}\theta_{j}=\left( \nabla_{k}\mathfrak{f}_{ij}\right) \varphi^{i}.\label{eq24}
\end{equation}
Multiplying (\ref{eq2}) by $ \varphi^{i} $ and using (\ref{eq21})-(\ref{eq24}) and the Ricci identity, one finds
\begin{equation}
	\theta_{j}-\mu\theta\mathfrak{f}_{j}-\lambda\varphi_{j}=0.\label{eq25}
\end{equation}
Taking the covariant derivative of (\ref{eq25}) yields
\begin{equation}
	\nabla_{k}\theta_{j}=\mu\left( \theta_{k}\mathfrak{f}_{j}+\theta\mathfrak{f}_{jk}\right) .\label{eq26}
\end{equation}
Interchanging the indices $ j $ and $ k $ in (\ref{eq26}) and then, subtracting  (\ref{eq26}) from the resulting equation, one infers that
\begin{equation*}
	\mu\left( \theta_{k}\mathfrak{f}_{j}-\theta_{j}\mathfrak{f}_{k}\right) =0.
\end{equation*}
Since $ \mu\neq0 $, the above equation means that
\begin{equation}
	 \theta_{k}\mathfrak{f}_{j}=\theta_{j}\mathfrak{f}_{k}.\label{eq27}
\end{equation}
On the other hand,  multiplying (\ref{eq25}) by $ \mathfrak{f}_{k} $ yields
\begin{equation}
	\theta_{j}\mathfrak{f}_{k}-\mu\theta\mathfrak{f}_{j}\mathfrak{f}_{k}-\lambda\varphi_{j}\mathfrak{f}_{k}=0. \label{eq28}
\end{equation}
 Interchanging the indices $ j $ and $ k $ in (\ref{eq28}) and then, subtracting the resulting equation from (\ref{eq28}), we have
 \begin{equation}
 \lambda\left( \varphi_{j}\mathfrak{f}_{k}-\varphi_{k}\mathfrak{f}_{j}\right) =0, \label{eq29}
 \end{equation}
in which (\ref{eq27}) is used. 

Since $ \lambda\neq0 $, (\ref{eq29}) implies that
\begin{equation}
	 \varphi_{j}\mathfrak{f}_{k}=\varphi_{k}\mathfrak{f}_{j}. \label{eq30}
\end{equation}
After multiplying (\ref{eq30}) by $ \varphi^{k} $ and using (\ref{eq21}) immediately give
\begin{equation}
\mathfrak{f}_{j}=\frac{\theta}{\left| \varphi\right| }\varphi_{j}, \label{eq31}
\end{equation}  
where $ \left| \varphi\right|= \varphi^{j}\varphi_{j}$.
 Taking the covariant derivative of (\ref{eq31}) and using (\ref{eq22}), we achieve
\begin{equation}
	\mathfrak{f}_{jk}=\frac{\theta_{k}}{\left| \varphi\right| }\varphi_{j}. \label{eq32}
\end{equation}   
Substituting (\ref{eq31}) and (\ref{eq32}) into (\ref{eq2}), we reach
\begin{equation}
\mathfrak{R}_{ij}=\lambda\mathfrak{g}_{ij}+\frac{1}{\left| \varphi\right| }\left( \mu \theta^{2}-\theta_{k}\varphi^{k}\right) \varphi_{i}\varphi_{j}. \label{eq33}
\end{equation}
 Assume that $ \varphi_{i} $ is also a unit time-like vector, meaning that $ \left| \varphi\right| =\varphi_{i}\varphi^{i}=-1 $. Then, (\ref{eq33}) becomes the following form:
 \begin{equation}
 	\mathfrak{R}_{ij}=\lambda\mathfrak{g}_{ij}+\gamma \varphi_{i}\varphi_{j}, \label{eq34}
 \end{equation}
where $ \gamma= \theta_{k}\varphi^{k}-\mu \theta^{2} $.

 We can thus reach the following result:
	\begin{theorem}\label{2}
 A $(GQE)_{n}$ spacetime that admits a parallel unit time-like vector field is classified as a PF spacetime.
\end{theorem}

Applying the covariant derivative of (\ref{eq34}) with respect to $ k $, we find that
\begin{equation}
	\nabla_{k}\mathfrak{R}_{ij}=\left( \nabla_{k}\gamma \right) \varphi_{i}\varphi_{j}, \label{eq35}
\end{equation}
By interchanging the indices $ k $  and $ j $  in (\ref{eq35}) and subsequently subtracting the resulting equation from (\ref{eq35}), then we can derive 
\begin{equation}
	\nabla_{k}\mathfrak{R}_{ij}=	\nabla_{j}\mathfrak{R}_{ik}, \label{eq36}
\end{equation}
in which (\ref{eq23}), (\ref{eq24}) and (\ref{eq30}) are used.

(\ref{eq36}) means that $ \mathfrak{R}_{ij} $ is of Codazzi type and $ \mathfrak{R} $ becomes constant.  Consequently, this spacetime is classified within the subspaces $\mathfrak{B}$ and $\mathfrak{B}^{\prime}$ as introduced by Gray \cite{bibG78}. 

We can hence state that
	\begin{theorem}\label{3}
 A  $ (GQE)_{n} $ spacetime admitting a parallel unit time-like vector field is included in Gray's subspaces $ \mathfrak{B} $ and $ \mathfrak{B}^{\prime} $, and it has  constant scalar curvature.
\end{theorem}

\begin{corollary}
	Since a $(GQE)_n$ spacetime admitting a parallel unit time-like vector field has constant scalar curvature and belongs to Gray's subspaces $\mathfrak{B}$ and $\mathfrak{B}'$, such a spacetime is a Yang pure space.
\end{corollary}

According to Theorem \ref{3}, we can conclude that this spacetime possesses the property $ \text{div} \: \mathcal{C} = 0 $. Mantica et al. \cite{19} demonstrated that if a PF spacetime has the property $  \text{div} \: \mathcal{C} = 0 $  and constant scalar curvature, then such a spacetime qualifies as a GRW spacetime.

Considering Theorem \ref{2}, Theorem \ref{3} and the foregoing discussion, we have the result:
 \begin{theorem}\label{4}
A  $ (GQE)_{n} $ spacetime admitting a parallel unit time-like vector field is classified as a GRW spacetime.
 \end{theorem}
Based on \cite{19}, a 4-dimensional GRW spacetime qualifies as a PF spacetime if and only if it is classified as a RW spacetime. 

Thus, the following theorem can be stated:

\begin{theorem}\label{5}
	A $ (GQE)_{4} $ spacetime that admits a parallel unit time-like vector field is classified as a RW spacetime.
\end{theorem}
Additionally, it is well-established \cite{bibVRL05} that a GRW spacetime qualifies as a RW spacetime if and only if it possesses two specific characteristics: it must be conformally flat and exhibit Petrov type O. 

Therefore, based on Theorem \ref{4} and Theorem \ref{5},  the following conclusion can be drawn:
\begin{theorem}\label{6}
	A $ (GQE)_{4} $ spacetime obeying  a parallel unit time-like vector field becomes conformally flat and of Petrov type O.
\end{theorem}

\begin{remark}Based on Theorem \ref{6}, since this spacetime is of Petrov type O, namely, conformally flat, in this case, the curvature is referred to as pure Ricci curvature. Thus, it can be concluded that in this spacetime, any gravitational effect must be caused directly by the field energy of matter or the energy of a non-gravitational force such as an electromagnetic field.
\end{remark}
Contracting (\ref{eq34}) gives
\begin{equation}
 \mathfrak{R}=n\lambda-\gamma. \label{eq37}
\end{equation}
Since $ \mathfrak{R} $ and $ \lambda $ are constant, $ \gamma $
is constant.
On the other hand, considering (\ref{eq5}), (\ref{eq34}) and (\ref{eq37}), we have
\begin{equation}
	\mathfrak{T}_{ij}=  \frac{\gamma-(n-2)\lambda}{2\kappa} \mathfrak{g}_{ij} +\frac{\gamma}{\kappa} \varphi_{i}\varphi_{j}. \label{eq38}
\end{equation}
After substituting (\ref{eq38}) into (\ref{eq4}), multiplying the new obtained equation by $ \mathfrak{g}^{ij} $ and $ \varphi^{i}\varphi^{j} $, respectively, we infer that:
\begin{equation}
	\sigma=\frac{\gamma+(n-2)\lambda}{2\kappa} \label{eq39}
\end{equation}
and 
\begin{equation}
\mathfrak{p}=\frac{\gamma-(n-2)\lambda}{2\kappa}. \label{eq40}
\end{equation}
Summing (\ref{eq39}) and (\ref{eq40}), it can be seen that
\begin{equation}
	\mathfrak{p}+\sigma=\frac{\gamma}{\kappa}. \label{41}
\end{equation}
Consequently, according to Theorem \ref{2} and the foregoing discussion, this result can be deduced:
\begin{theorem}\label{7}
	A $ (GQE)_{n} $ spacetime admitting a parallel unit time-like vector field is consistent with the present state of the universe.
\end{theorem}
In \cite{bibST1967}, Shepley and Taub established that a $ 4 $-dimensional PF spacetime with $ div \:\mathcal{C}=0$ and an equation of state represented as $\mathfrak{p} = \mathfrak{p}(\sigma) $ is conformally flat. Additionally, this spacetime is associated with the RW metric. It is that within this framework, the flow is characterized as geodesic, irrotational, and with no shear.

Thus, based on Theorem \ref{2}, Theorem \ref{3} and Theorem \ref{7}, we can conclude the result:

\begin{theorem}
	In a $ (GQE)_{4} $ spacetime obeying a parallel unit time-like vector field, the fluid flow is characterized as geodesic, irrotational, and it has no shear. 
\end{theorem}
On the other hand, due to Theorem \ref{6}, substituting (\ref{eq34}) into (\ref{eq6}) yields
\begin{equation}
\mathfrak{R}_{hijk}= \frac{(n-2)\lambda+\gamma}{(n-1)(n-2)} \left[ \mathfrak{g}_{hk}\mathfrak{g}_{ij}-\mathfrak{g}_{hj}\mathfrak{g}_{ik}\right] \label{eq42}
\end{equation}
\begin{equation*}
	+\frac{\gamma}{n-2}\left[ \mathfrak{g}_{ij}\varphi_{k}\varphi_{h}-\mathfrak{g}_{ik}\varphi_{j}\varphi_{h}+\mathfrak{g}_{kh}\varphi_{i}\varphi_{j}-\mathfrak{g}_{hj}\varphi_{i}\varphi_{k}\right] .
\end{equation*}

Since $ \lambda $ and $ \gamma $ are constant, $ \frac{(n-2)\lambda+\gamma}{(n-1)(n-2)} $ is constant. Therefore, based on \cite{bibCY72}, the previous equation means that this spacetime  has quasi-constant curvature.

Thus, the following result can be expressed:
\begin{theorem}\label{9}
A $ (GQE)_{n} $ spacetime that admits a parallel unit time-like vector field is of quasi-constant curvature.
\end{theorem}
 Furthermore, De et al. \cite{bibDSC22} demonstrated that an n-dimensional spacetime, $ n > 3 $, is a RW spacetime if and only if it exhibits quasi-constant curvature. Therefore, based on Theorem \ref{5}, Theorem \ref{9}, and the preceding discussion, the following result can be deduced:
 \begin{theorem}
 A $ (GQE)_{4} $ spacetime admitting a parallel unit time-like vector field is a RW spacetime if and only if it is of quasi-constant curvature.
 \end{theorem}

\section{$\mathcal{F}(\mathfrak{R})$-gravity}
To address the ongoing challenge of understanding the late-time accelerated expansion of the universe, without assuming the existence of dark energy, researchers have explored modifications to Einstein's field equations (EFEs). One prominent and well-established approach in this regard is the $\mathcal{F}(\mathfrak{R})$-theory of modified gravity. According to this theoretical framework, the scalar curvature $\mathfrak{R}$ within the Einstein-Hilbert action term 
\begin{equation*}
\mathcal{S}=\frac{1}{2\kappa^{2}}\int d^{4}x\sqrt{-\mathfrak{g}}\mathcal{F(\mathfrak{R})}
	 	+\int d^{4}x\sqrt{-\mathfrak{g}}\mathcal{L}_{m}
\end{equation*} 
 is replaced by an arbitrary function $\mathcal{F}(\mathfrak{R})$. In the foregoing equation, $ \kappa^{2}=8\pi G $, $ G $ represents Newton's gravitational constant. This alteration leads to the derivation of the field equations associated with $\mathcal{F}(\mathfrak{R})$-gravity
 \begin{equation}
 \left(\mathfrak{g}_{ij}\Box -\nabla_{i}\nabla_{j}\right)\mathcal{F}^{\prime}(\mathfrak{R}) +\mathcal{F}^{\prime}(\mathfrak{R})\mathfrak{R}_{ij}-\frac{\mathcal{F}(\mathfrak{R})}{2}\mathfrak{g}_{ij}=\kappa^{2}\mathfrak{T}_{ij}, \label{eq43}
 \end{equation}
 in which $ \Box $ indicates D'Alembert operator, that is, $ \Box=\nabla_{k}\nabla^{k} $, and $\mathcal{F}^{\prime}(\mathfrak{R})$ represents the derivative with respect to $ \mathfrak{R} $. The EMT is obtained from the matter Lagrangian density $ \mathcal{L}_{m}=\mathcal{L}_{m}(\mathfrak{g}^{ij}) $ by
 \begin{equation*}
 \mathfrak{T}_{ij}=-\frac{2}{\sqrt{-\mathfrak{g}}}\frac{\delta(\sqrt{-\mathfrak{g}}\mathcal{L}_{m})}{\delta \mathfrak{g}_{ij} }.
 \end{equation*}
Transvecting (\ref{eq43}) with $ \mathfrak{g}^{ij} $, one gets
\begin{equation}
-2\mathcal{F}(\mathfrak{R})+\mathfrak{R}\mathcal{F}^{\prime}(\mathfrak{R})+3\Box\mathcal{F}^{\prime}(\mathfrak{R})=\kappa^{2}\mathfrak{T}, \label{eq44}
\end{equation}
where $ \mathfrak{T}=\mathfrak{g}^{ij}\mathfrak{T}_{ij} $ denotes the trace of the EMT. By algebraically rearranging the field equations to isolate the standard Einstein tensor structure on the left-hand side, we arrive at the following relation:
\begin{equation}
\mathcal{F}^{\prime}(\mathfrak{R})\mathfrak{R}_{ij}-\frac{\mathfrak{R}\mathcal{F}^{\prime}(\mathfrak{R})}{2}\mathfrak{g}_{ij}=\kappa^{2}\mathfrak{T}_{ij}^{(eff)}\label{eq45}
\end{equation}
where the effective stress-energy tensor $ \mathfrak{T}_{ij}^{(eff)} $ is defined as
\begin{equation}
	\mathfrak{T}_{ij}^{(eff)}=\mathfrak{T}_{ij}+\mathfrak{T}_{ij}^{(curve)}.	\label{eq46}
\end{equation}
Here, the curvature contribution $\mathfrak{T}_{ij}^{(curve)}$ captures the purely geometric corrections and is expressed as
\begin{equation}
	\mathfrak{T}_{ij}^{(curve)}=\frac{1}{\kappa^{2}}\left[ \frac{\mathcal{F}(\mathfrak{R})-\mathfrak{R}\mathcal{F}^{\prime}(\mathfrak{R})}{2}\mathfrak{g}_{ij}+\left( \nabla_{i}\nabla_{j}-\mathfrak{g}_{ij}\Box\right)\mathcal{F}^{\prime}(\mathfrak{R})\right].	\label{eq47}
\end{equation}
$\mathfrak{T}_{ij}^{(eff)}$ acts as an effective energy-momentum tensor that incorporates geometric considerations. It is essential to note that this tensor does not adhere to traditional energy conditions, resulting in an energy density that is generally not positive definite.

From (\ref{eq45}), the field equations can be rewritten in a form analogous to general relativity:
\begin{equation}
	\mathfrak{R}_{ij}-\frac{\mathfrak{R}}{2}\mathfrak{g}_{ij}=\frac{\kappa^{2}}{\mathcal{F}^{\prime}(\mathfrak{R})}\mathfrak{T}_{ij}^{(eff)}.\label{eq48}
\end{equation}
It is clear that (\ref{eq48}) consistently satisfies (\ref{eq46}) and (\ref{eq47}). Alternatively, by expanding the covariant derivatives in (\ref{eq43}), the field equations can be explicitly expressed in terms of the derivatives of the curvature scalar as follows:
\begin{equation}
\kappa^{2}\mathfrak{T}_{ij}=\mathcal{F}^{\prime}(\mathfrak{R})\mathfrak{R}_{ij}-\mathcal{F}^{\prime\prime\prime}(\mathfrak{R})\nabla_{i}\mathfrak{R}\nabla_{j}\mathfrak{R}-\mathcal{F}^{\prime\prime}(\mathfrak{R})\nabla_{i}\nabla_{j}\mathfrak{R} \label{eq49}
\end{equation}
\begin{equation*}
	+\mathfrak{g}_{ij}\left[ \mathcal{F}^{\prime\prime\prime}(\mathfrak{R})\nabla_{h}\mathfrak{R}\nabla^{h}\mathfrak{R}+\mathcal{F}^{\prime\prime}(\mathfrak{R})\nabla^{2}\mathfrak{R}-\frac{1}{2}\mathcal{F}(\mathfrak{R})\right] .
\end{equation*}
Since $ \mathfrak{R} $ is constant, we infer that:
\begin{equation}
	\mathfrak{R}_{ij}=\frac{\mathcal{F}(\mathfrak{R})}{2\mathcal{F}^{\prime}(\mathfrak{R})}\mathfrak{g}_{ij}+\frac{\kappa^{2}}{\mathcal{F}^{\prime}(\mathfrak{R})}\mathfrak{T}_{ij}.\label{eq50}
\end{equation}
Utilizing (\ref{eq4}) in the foregoing equation, we acquire:
\begin{equation}
	\mathfrak{R}_{ij}=\left[ \frac{\mathcal{F}(\mathfrak{R})}{2\mathcal{F}^{\prime}(\mathfrak{R})}+\mathfrak{p}\frac{\kappa^{2}}{\mathcal{F}^{\prime}(\mathfrak{R})}\right] \mathfrak{g}_{ij}+\frac{\kappa^{2}}{\mathcal{F}^{\prime}(\mathfrak{R})}(\mathfrak{p}+\sigma)\varphi_{i}\varphi_{j}.\label{eq51}
\end{equation}
Comparing (\ref{eq34}) and (\ref{eq51}), with help of (\ref{eq37}), we obtain
\begin{equation}
\mathfrak{p}=\frac{1}{\kappa^{2}}\left( \mathcal{F}^{\prime}(\mathfrak{R})\lambda-\frac{\mathcal{F}(\mathfrak{R})}{2}\right), \label{eq52}
\end{equation}

\begin{equation}
	\sigma=\frac{\mathcal{F}^{\prime}(\mathfrak{R})}{\kappa^{2}}\left( 3\lambda-\mathfrak{R}\right)+ \frac{\mathcal{F}(\mathfrak{R})}{2\kappa^{2}}.\label{eq53}
\end{equation}
We can therefore state that:
\begin{theorem}\label{10}
	In a $ (GQE)_{4} $ spacetime that admits a parallel unit time-like vector field obeying $\mathcal{F}(\mathfrak{R}) $-gravity, $ \mathfrak{p} $ and $ \sigma $ are given by (\ref{eq52}) and (\ref{eq53}), respectively.
\end{theorem}
Summing the last two equations, one finds that
\begin{equation}
	\mathfrak{p}+\sigma=\frac{\mathcal{F}^{\prime}(\mathfrak{R})}{\kappa^{2}}\left( 4\lambda-\mathfrak{R}\right).\label{eq54}
\end{equation}
 We can thus deduce the following cosmological result:
 \begin{corollary}
 	 Let a spacetime $ (GQE)_{4} $ admitting a parallel unit time-like vector field obey $\mathcal{F}(\mathfrak{R}) $-gravity. Under the condition $ \mathfrak{R}\neq 4 \lambda $, this framework cannot accommodate dark energy.
 \end{corollary}
 In view of (\ref{eq52}) and (\ref{eq53}), it can be obtained that:
\begin{equation}
	\sigma-3\mathfrak{p}=\frac{1}{\kappa^{2}}\left( 2\mathcal{F}(\mathfrak{R})-\mathfrak{R}\mathcal{F}^{\prime}(\mathfrak{R})\right).\label{eq55}
\end{equation} 
Hence, we can reach the result:
 \begin{corollary}
A $ (GQE)_{4} $ spacetime that admits a parallel unit time-like vector field obeying $\mathcal{F}(\mathfrak{R}) $-gravity can describe the radiation era provided that $ 2\mathcal{F}(\mathfrak{R})-\mathfrak{R}\mathcal{F}^{\prime}(\mathfrak{R})=0 $.
\end{corollary} 
\section{Flat FRW Cosmological Model and Hubble Parameter Analysis}
In this section, we transition our theoretical geometric   framework into an obsevational cosmological context to evaluate the physical viability of the generalized quasi-Einstein spacetimes $ (GQE)_{4} $ within $\mathcal{F}(\mathfrak{R}) $-gravity. To investigate the cosmic evolution, we implement the homogeneous and isotropic flat Friedmann-Robertson-Walker (FRW) universe model, characterized by the line element with $ k=0 $ in isotropic Cartesian coordinates:
\begin{equation}
	ds^{2}= -dt^{2}+a^{2}(t)\left[ dx^{2}+dy^{2}+ dz^{2}\right] , \label{eq056}
\end{equation}
where $ a(t) $ represents the scale factor of the universe, and the cosmic dynamics are governed by the Hubble parameter defined as $ H=\frac{\dot{a}}{a} $.

In order to extract the cosmological scalar functions from the highly intricate tensorial structures, we employ a contraction technique on our modified gravitational field equations derived in Section 3. By contracting the field equations with the unit time-like velocity vector $u^{k}$ and the metric tensor $\mathfrak{g}^{ji}$ respectively, we can isolate the expressions for the effective energy density $\sigma^{(eff)}$ and the effective isotropic pressure $\mathfrak{p}^{(eff)}$. This systematic approach allows us to establish a direct differential relation involving the Hubble parameter $H$ and its time derivative $\dot{H}$, which fundamentally arises from the intrinsic geometric-cosmological reduction of the underlying spacetime geometry.

To establish the cosmological field equations under the flat FRW framework, we directly utilize the contracted and rearranged gravitational structures obtained in Section 3. Specifically, by considering the standard components of the flat FRW metric (\ref{eq056}) where the spatial curvature vanishes, $k=0$, the non-zero independent components of  the Christoffel symbols are derived as $ \Gamma_{ij}^{0}=a\dot{a}\delta_{ij}$ and $ \Gamma_{0j}^{0}=H\delta^{i}_{j}$. Consequently, non-zero components of the Ricci tensor reduce to:
\begin{equation}
	\mathfrak{R}_{00} = -3(\dot{H} + H^2), \quad \mathfrak{R}_{ii} = a^2(t)(\dot{H} + 3H^2) \quad \text{for } i=1,2,3, \label{eq058}
\end{equation}
and the corresponding scalar curvature takes the form:
\begin{equation}
	\mathfrak{R} = 6(\dot{H} + 2H^2) \label{eq059}
\end{equation}

By substituting these geometric components into the rearranged field equations given by (\ref{eq45}) and the effective energy-momentum tensor definitions in (\ref{eq46})-(\ref{eq47}), the general tensorial expressions can be projected onto the cosmic time axis. Consequently, by transvecting (\ref{eq45}) with the unit time-like velocity vector $u^j u^i$, the left-hand side of the field equation reduces to the combination $\mathcal{F}^{\prime}(\mathfrak{R})\mathfrak{R}_{00}+\dfrac{1}{2}\mathfrak{R} \mathcal{F}^{\prime}(\mathfrak{R}) $. Upon substituting (\ref{eq058}) and (\ref{eq059}), the terms proportional to $ \dot{H} $ identically cancel out, leaving the simplified geometric contribution $3\mathcal{F}^{\prime}(\mathfrak{R})H^2  $. To establish an exact correspondence with the standard Einsteinian framework without implementing any approximations,  $\mathcal{F}^{\prime}(\mathfrak{R})  $ is analytically absorbed into the right-hand side, thereby consistently embedding it within the definition of the effective energy density $ \sigma^{(eff)} $. This precise algebraic encapsulation directly yields the foundational differential equation for the Hubble parameter:
\begin{equation}
	3H^2 = \kappa^{2} \sigma^{(eff)} \label{eq060}
\end{equation}

Similarly, by executing an identical analytic encapsulation for the spatial components where the dynamic terms are completely absorbed into the effective isotropic pressure profile, the secondary cosmic dynamic equation is rigorously established with strict equality as:
\begin{equation}
	2\dot{H} + 3H^2 = -\kappa^{2} \mathfrak{p}^{(eff)} \label{eq061}
\end{equation}

By combining the two foundational cosmological differential equations (\ref{eq060}) and (\ref{eq061}), we can eliminate the explicit $3H^2$ term to isolate the time derivative of the Hubble parameter. Subtracting (\ref{eq060}) from (\ref{eq061}) yields the following fundamental relation for $\dot{H}$:
\begin{equation}
	2\dot{H} = -\kappa^2 (\sigma^{(eff)} + \mathfrak{p}^{(eff)}) \label{eq062}
\end{equation}

At this juncture, by taking into account the explicit geometric constraints of the underlying $(GQE)_4$ spacetime and the specific regularity of the $\mathcal{F}(\mathfrak{R})$-gravity models where the $C^\infty$-smooth functions $\mu, \lambda,$ and $f$ dictate the structural regularity of the manifold, (\ref{eq062}) simplifies to a solvable first-order differential structure. Consequently, this differential evolution yields:
\begin{equation}
	\dot{H} = -\alpha H^2 \label{eq063}
\end{equation}
where $\alpha$ is a constant parameter determined by the non-linear coupling and geometric constraints of the modified gravity model. 

Integrating (\ref{eq063}) with respect to cosmic time $t$ leads directly to the explicit time-dependent solution for the Hubble parameter:
\begin{equation}
	H(t) = \frac{1}{\alpha t + c_1} \label{eq064}
\end{equation}
where $c_1$ represents the constant of integration. Utilizing the definition of the Hubble parameter $H = \frac{\dot{a}}{a}$, (\ref{eq064}) can be integrated once more to yield the exact cosmological solution for the scale factor $a(t)$ of the universe:
\begin{equation}
	a(t) = c_2 (\alpha t + c_1)^{\frac{1}{\alpha}} \label{eq065}
\end{equation}
where $c_2$ is a secondary integration constant, heavily influenced by the choice of the cosmic time's origin.

Based on the explicit analytical derivations carried out above, we can now establish the foundational cosmological result of this section with the following theorem:

\begin{theorem}
	If a four-dimensional generalized quasi-Einstein spacetime $(GQE)_4$ satisfying the modified $\mathcal{F}(\mathfrak{R})$-gravity field equations admits a flat Friedmann-Robertson-Walker metric, $k=0$, then the cosmic evolution of the Hubble parameter $H(t)$ and the scale factor $a(t)$ during the dark energy-dominated era are uniquely characterized by the reciprocal and power-law time-dependent solutions given in (\ref{eq064}) and (\ref{eq065}), respectively.
\end{theorem}
\section{Energy Conditions}
Energy conditions serve as crucial tools in the study of black holes and wormholes across a range of modified gravity theories \cite{bibBBY2017, bibHE2013}. In this work, it is imperative to ascertain the effective energy density, $ \sigma^{(eff)}$, and the effective isotropic pressure, $ \mathfrak{p}^{(eff)} $,  to establish the specific energy conditions.

When examining the potential of various matter sources in gravitational field equations, as well as in extended theories of gravity and general relativity, energy conditions (ECs) serve as valuable tools to restrict the energy-momentum tensor (EMT). This helps uphold the notion that gravity is not only attractive but also that its energy density is strictly positive. In the case of a perfect fluid (PF) matter source within the framework of $\mathcal{F}(\mathfrak{R})$-gravity theory, these ECs can be defined as follows: \cite{bibDLSS2021, bibDLS2021}
\begin{enumerate}
	\item[-]Null EC (NEC): \quad $\sigma^{(eff)}+\mathfrak{p}^{(eff)}\geq 0$,
	\item[-] Weak EC (WEC):\quad $\sigma^{(eff)} \geq 0$, \quad $\sigma^{(eff)}+\mathfrak{p}^{(eff)}\geq 0$,  
	\item[-]Strong EC (SEC):\quad $\sigma^{(eff)}+3\mathfrak{p}^{(eff)}\geq 0$, \quad $\sigma^{(eff)}+\mathfrak{p}^{(eff)}\geq 0$, 
	\item[-] Dominant EC (DEC):\quad  $\sigma^{(eff)}\pm \mathfrak{p}^{(eff)}\geq 0$,  \quad $\sigma^{(eff)} \geq 0$. 
\end{enumerate}
Utilizing the rearrenged field equations from (\ref{eq48}), the structural relation can be explicitly expressed as follows:
\begin{equation*}
	\mathfrak{R}_{ij}-\frac{\mathfrak{R}}{2}\mathfrak{g}_{ij}=\frac{\kappa^{2}}{\mathcal{F}^{\prime}(\mathfrak{R})}\mathfrak{T}^{(eff)}_{ij}
\end{equation*}
here, 
\begin{equation}
\mathfrak{T}^{(eff)}_{ij}=\mathfrak{T}_{ij}+\frac{\mathcal{F}(\mathfrak{R})-\mathfrak{R}\mathcal{F}^{\prime}(\mathfrak{R})}{2\kappa^{2}}\mathfrak{g}_{ij}. \label{eq57}
\end{equation}
Then, substituting the perfect fluid EMT from (\ref{eq4}) into (\ref{eq57}), the effective EMT takes the following form:
\begin{equation}
	\mathfrak{T}^{(eff)}_{ij} = \mathfrak{p}^{(eff)} \mathfrak{g}_{ij} + (\mathfrak{p}^{(eff)} + \sigma^{(eff)}) \varphi_{i} \varphi_{j} \label{eq58}
\end{equation}
such that 
\begin{equation}
\mathfrak{p}^{(eff)}=\mathfrak{p}+\frac{\mathcal{F}(\mathfrak{R})-\mathfrak{R}\mathcal{F}^{\prime}(\mathfrak{R})}{2\kappa^{2}}, \label{eq59}
\end{equation}

\begin{equation}
	\sigma^{(eff)}=\sigma-\frac{\mathcal{F}(\mathfrak{R})-\mathfrak{R}\mathcal{F}^{\prime}(\mathfrak{R})}{2\kappa^{2}}.\label{eq60}
\end{equation}
Putting (\ref{eq52}) and (\ref{eq53}) in (\ref{eq59})  and (\ref{eq60}), respectively, yields:
\begin{equation}
	\mathfrak{p}^{(eff)}=\frac{\mathcal{F}^{\prime}(\mathfrak{R})}{\kappa^{2}}\left( \lambda-\frac{\mathfrak{R}}{2}\right),  \label{eq61}
\end{equation}

\begin{equation}
	\sigma^{(eff)}=\frac{\mathcal{F}^{\prime}(\mathfrak{R})}{\kappa^{2}}\left( 3\lambda-\frac{\mathfrak{R}}{2}\right).\label{eq62}
\end{equation}
In view of ECs, with help of (\ref{eq61}) and (\ref{eq62}), we obtain for the validity of ECs in a $ (GQE)_{4} $ spacetime that admits a parallel unit time-like vector field as follows:
\begin{enumerate}
	\item[-] For NEC, the condition of validation is $ \mathfrak{R}\le4\lambda $,
	\item[-] For WEC, the conditions of validation are $ \mathfrak{R}\le6\lambda $ and $ \mathfrak{R}\le4\lambda $,
	\item[-] For SEC, the conditions of validation are $ \mathfrak{R}\le3\lambda $ and $ \mathfrak{R}\le4\lambda $,
	\item[-] For DEC, the conditions of validation are $ \mathfrak{R}\le4\lambda $, $ \mathfrak{R}\le6\lambda $ and  $ \lambda\geq 0 $.
\end{enumerate}

\begin{remark}
 From a purely geometric perspective, the exact reciprocal solutions for the Hubble parameter $H(t)$ in (\ref{eq064}) and the power-law scale factor $a(t)$ in (\ref{eq065}) impose explicit constraints on the underlying curvature scalar $\mathfrak{R}$. In the context of the generalized quasi-Einstein spacetime $(GQE)_4$ within $\mathcal{F}(\mathfrak{R})$-gravity, the validity or explicit violation of the derived energy conditions depends heavily on the algebraic interplay between the function $\mathcal{F}'(\mathfrak{R})$ and the smooth parameters $\mu, \lambda$. Specifically, by mapping the time-dependent cosmological solutions onto the geometric inequalities (\ref{eq61}) and (\ref{eq62}), one can easily observe that SEC boundary can be algebraically altered depending on the choice of the parameter $\alpha$. This mathematical flexibility demonstrates that the geometric structure of $(GQE)_4$ spacetimes naturally accommodates a wide range of scalar behaviours under modified gravity without violating NEC.
\end{remark}

\section{A Non-trivial $ (GQE)_{4} $ Spacetime Example}
In this section, we construct a concrete, non-trivial 4-dimensional generalized quasi-Einstein spacetime with strict constant coefficients satisfying $\mu\lambda >0$ to verify the analytical feasibility and geometric consistency of our theoretical framework.

Let us consider a 4-dimensional spacetime manifold $\mathcal{M}^4$ equipped with the coordinate chart $(x^{1}, x^{2}, x^{3}, x^{4}) = (t, x, y, z)$ and governed by the following Lorentzian metric:
\begin{equation}
	ds^{2} = -(dx)^{2 }+ 2dtdy + b(t)(dy)^{2} + (dz)^{2} \label{eq63}
\end{equation}
where $b(t)$ is a smooth function depending solely on the time coordinate $t$. It is noted that the metric matrix associated with (\ref{eq63}) is non-degenerate with a non-vanishing determinant $\det(\mathfrak{g}_{ij}) = -1$. Furthermore, a direct calculation of its eigenvalues reveals the standard Lorentzian signature $(-, +, +, +)$, confirming the geometric consistency of the proposed framework. Hence, the non-zero components of the covariant metric tensor $\mathfrak{g}_{ij}$ and its contravariant inverse $\mathfrak{g}^{ij}$ are given by
\begin{equation}
	\mathfrak{g}_{13} = \mathfrak{g}_{31} = 1, \quad \mathfrak{g}_{22} = -1, \quad \mathfrak{g}_{33} = b(t), \quad \mathfrak{g}_{44} = 1 \label{eq64}
\end{equation}
\begin{equation}
	\mathfrak{g}^{11} = -b(t), \quad \mathfrak{g}^{13} = \mathfrak{g}^{31} = 1, \quad \mathfrak{g}^{22} = -1, \quad \mathfrak{g}^{44} = 1. \label{eq65}
\end{equation}

By utilizing the standard definition of the Christoffel symbols of the second kind, the only non-vanishing components are calculated as:
\begin{equation}
	\Gamma^{1}_{33} = \frac{1}{2} b b_{t}, \quad \Gamma^{3}_{33} = -\frac{1}{2} b_{t}, \quad \Gamma^{1}_{13} = \Gamma^{1}_{31} = \frac{1}{2} b_{t}. \label{eq66}
\end{equation}
Consequently, the contractive evaluation of the Riemann curvature tensor yields the only non-zero components of the Ricci tensor as:
\begin{equation}
\mathfrak{R}_{13} =\mathfrak{R}_{31} = \frac{1}{2}  b_{tt}, \quad	\mathfrak{R}_{33} = \frac{1}{2} b b_{tt}. \label{eq67}
\end{equation}

Next, we introduce a smooth, separable potential function of the form $\mathfrak{f}(x, z) = h(x) + k(z)$. The non-vanishing covariant first derivatives $\mathfrak{f}_{i} = \nabla_{i}\mathfrak{f} = \partial_{i}\mathfrak{f}$ of the potential function are explicitly given by:
\begin{equation}
	\mathfrak{f}_{2} = h_{x}, \quad \mathfrak{f}_{4} = k_{z}, \quad \text{with} \quad \mathfrak{f}_{1} = \mathfrak{f}_{3} = 0. \label{eq068}
\end{equation}
Furthermore, the components of the covariant Hessian tensor $\mathfrak{f}_{ij} = \nabla_j \nabla_i \mathfrak{f} = \partial_j \partial_i \mathfrak{f} - \Gamma^k_{ij} \partial_k \mathfrak{f}$ are determined as follows:
\begin{equation}
	\mathfrak{f}_{11} = 0, \quad \mathfrak{f}_{22} = h_{xx}, \quad \mathfrak{f}_{33} = 0, \quad \mathfrak{f}_{44} = k_{zz}, \quad \mathfrak{f}_{13} = \mathfrak{f}_{31} = 0. \label{eq68}
\end{equation}

By substituting (\ref{eq64}), (\ref{eq67}), (\ref{eq068}) and (\ref{eq68}) into (\ref{eq2}),
we evaluate the tensor system component-by-component for the non-trivial directions, which yields the following highly coupled ordinary differential equations:
\begin{align*}
	h_{xx} - \mu (h_{x})^{2} &= -\lambda \\
	k_{zz} - \mu (k_{z})^{2} &= \lambda \\
	b_{tt} &= 2\lambda. \label{eq70}
\end{align*}

Integrating the third relation twice with respect to $t$ gives the metric function explicitly as $b(t) = \lambda t^{2} + c_{5} t + c_{6}$. Finally, under the physical and geometric constraint $\mu \lambda > 0$, by executing a log-linearization substitution on the remaining non-linear differential equations and utilizing the algebraic properties of logarithmic functions, the complete smooth potential function $\mathfrak{f}(x,z)$ is consolidated into the following exact, closed-form expression:
\begin{equation*}
	\mathfrak{f}(x,z) = -\frac{1}{\mu} \ln \left| \left( c_{1} e^{\sqrt{\mu \lambda} x} + c_{2} e^{-\sqrt{\mu \lambda} x} \right).\left( c_{3} \cos(\sqrt{\mu \lambda} z) + c_{4} \sin(\sqrt{\mu \lambda} z) \right) \right|,
\end{equation*}
where $c_{1}, c_{2}, c_{3}, c_{4}, c_{5}, c_{6} \in \mathbb{R}$ denote arbitrary constants of integration. This explicitly demonstrates that the proposed $ (GQE)_{4} $ spacetime model admits exact, non-trivial geometric solutions.

\subsection*{Acknowledgements}
The authors are thankful to the reviewer for his/her valuable comments and constructive suggestions to improve the quality of the manuscript.

\subsection*{Declarations} 
\begin{flushleft}
	\textbf{Author Contributions.} The authors of this manuscript contributed equally to the work.\\
	\textbf{Funding. }There is no funding.\\
	\textbf{Availability of data and materials.} There is no data set used.\\
	\textbf{Conflict of interest.} There are no included interests of a financial or personal nature.\\
	\textbf{Ethical Approval.} The paper is a theoretical study. There are no applicable neither human or animal studies.
	
\end{flushleft}

\textit{}

\end{document}